# Do Wilson Fermions Induce an Adjoint Gauge Action?

L. Kärkkäinen[a] and K. Rummukainen[b] *,

[a]Department of Physics, University of Arizona, Tucson, AZ 85721, USA

[b]Department of Physics, Indiana University, Bloomington, IN 47405, USA

Expansions of the Wilson determinant in lattice QCD with quarks produce gauge action terms which shift the coupling constant of the fundamental representation plaquette action and induce an adjoint representation plaquette action. We study the magnitude of these induced couplings with two flavors of Wilson fermions. We utilize a microcanonical demon method, which allows us to measure the induced couplings directly from gauge configurations generated by full fermionic simulations.

hep-lat/9411025  15 Nov 94

## 1. Introduction

Lattice QCD simulations with two light Kogut-Susskind quarks seem to indicate that QCD with two massless quarks has a second order finite temperature deconfining phase transition, which is transformed into a smooth crossover if the quarks are massive. This scenario is also consistent with the universality arguments [1]. However, with two flavors of Wilson quarks recent simulations have shown that at large values of the hopping parameter the phase diagram is more complicated [2]: with $N_t = 6$ and at hopping parameter $\kappa \sim 0.19$ the system has a strong first order like transition, where the plaquette expectation value has a sharp discontinuous jump. However, the average Polyakov loop remains small at this point, and starts to increase only when $\beta$ is considerably larger. This behaviour is strongly reminiscent of separated bulk and thermal transitions; this occurs, for example, in pure gauge SU(3) simulations with mixed fundamental – adjoint action [3]. In fact, these systems resemble each other so much that one is tempted to assume that the *dynamical* reason for these transitions is the same; i.e. that the Wilson fermionic action induces a strong adjoint pure gauge action.

The pure gauge SU(3) fundamental-adjoint action can be written as $S = \beta_F S_F + \beta_A S_A$, where

$$S_F = \sum_P (1 - \frac{1}{3}\operatorname{Re}\operatorname{Tr} U_P) \quad (1)$$

$$S_A = \sum_P (1 - \frac{1}{9}|\operatorname{Tr} U_P|^2) \quad (2)$$

where $\operatorname{Tr} U_P$ is the fundamental representation trace. The gauge action induced by the fermion determinant can be studied perturbatively via the hopping parameter expansion; the leading terms are $\Delta\beta_F \propto \kappa^4$ and $\Delta\beta_A \propto \kappa^{12}$. This expansion is accurate only when $\kappa$ is very small. Results with much broader validity range can be obtained with the heavy quark perturbation. This method was used by Hasenfratz and DeGrand [7] to calculate $\Delta\beta_F$, which agrees very well with the MC data. However, this expansion has not been performed for $\Delta\beta_A$.

## 2. Demon Algorithm

We use the microcanonical demon method [4–6] to project out the induced gauge action from full fermionic simulations. Our ansatz for the effective action is given by eqs. (1,2), with a priori unknown coupling constants $\beta_i = \beta_F$ or $\beta_A$. For both of the coupling constants we introduce a *demon*, which is a real-valued action variable with $0 \le D_i \le D_i^{\max}$. Effectively, the fermionic action is used as a heat bath to thermalize the demons: a configuration generated with the original action is updated microcanonically with the demons keeping $S_i + D_i$ constant for all $i$ separately. At the end of the microcanonical update, we discard the old configuration and substitute it with a new one, while preserving the values of the demons. Repeating this many times the demons attain an equilibrium distribution and we

---

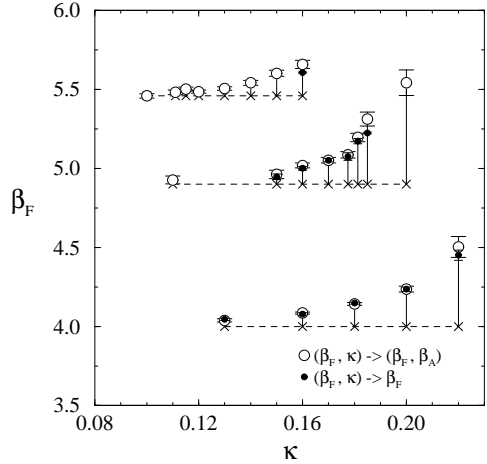

Figure 1. The induced $\beta_F$. Small crosses on horizontal dashed lines indicate the simulation $(\beta_F^0, \kappa)$, opaque circles and black dots the measured $\beta_F$-values using the effective fundamental-adjoint and fundamental only gauge action, respectively. The length of the vertical bars gives the magnitude of the induced $\Delta\beta_F$.

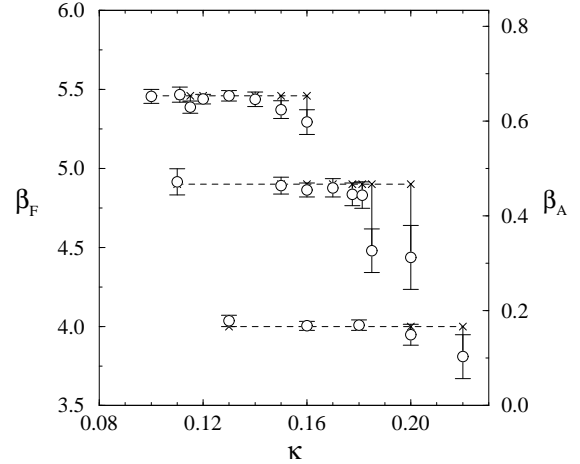

Figure 2. The induced $\beta_A$. The scale on the right gives the magnitude of the induced $\beta_A$, measured as the vertical distance between the plot symbols and dashed horizontal lines. Points below the dashed lines indicate negative $\beta_A$.

can measure $\langle D_i \rangle$. The induced $\beta_i$ can be solved from the equation

$$\langle D_i \rangle = \frac{1}{\beta_i} - \frac{D_i^{\max}}{\exp(\beta_i D_i^{\max}) - 1}. \qquad (3)$$

In our case, only the adjoint action demon was bounded from above ($D_F^{\max} = \infty$).

There are various non-equivalent ways to perform the demon update. For each starting configuration, one can update microcanonically until the demons and the system are properly thermalized, or one can stop the microcanonical update after one – or even partial – update sweep. If the effective action is completely equivalent to the original action, these two methods give the same unique answer. However, in this case we have a dramatically truncated effective action, and these method need not be equivalent. Nevertheless, the differences are minor if the heat capacity of the demons is much smaller than that of the system, and in all our tests the possible differences were completely overwhelmed by the statistical noise.

## 3. Simulations and Results

We tested the demon method by applying it to pure gauge fundamental-adjoint simulation. With $4^4$-lattice, we simulated the system at couplings $(\beta_F, \beta_A)_0 = (3.6, 1.8)$ and $(4.0, 2.0)$. The induced couplings were $(3.594(6), 1.812(17))$ and $(4.017(27), 1.92(6))$, respectively, compatible with the input values. The latter coupling pair is very close to the bulk transition line in $(\beta_F, \beta_A)$-plane. When we used an effective action consisting only of the $S_F$-part, $\beta_F$ was $4.688(9)$ and $6.044(12)$; the latter value is very close to the extension of the bulk transition line to the $\beta_A = 0$ -axis [3].

If the Wilson fermion action induces a strong adjoint coupling giving rise to bulk fundamental-adjoint transition, one should be able to observe the induced couplings already in small volumes. We performed simulations on $4^4$ lattices with 28 different $(\beta_F^0, \kappa)$ pairs. In figs. 1 and 2 we show the measured $\beta_F$ and $\beta_A$ calculated with $\beta_F^0 = 4.0, 4.9, 5.46$, and several $\kappa$ values; in fig. 3 with constant $\kappa = 0.19$. In figs. 1 and 3 we also show $\beta_F$ when $S_{\text{EFF}} = \beta_F S_F$ only.

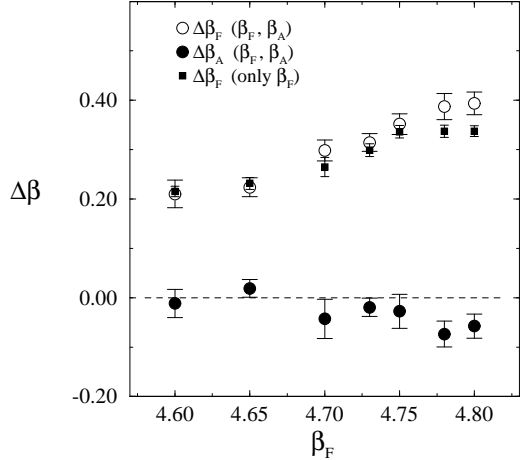

Figure 3. The induced $\beta_F$ and $\beta_A$, when $\kappa = 0.19$.

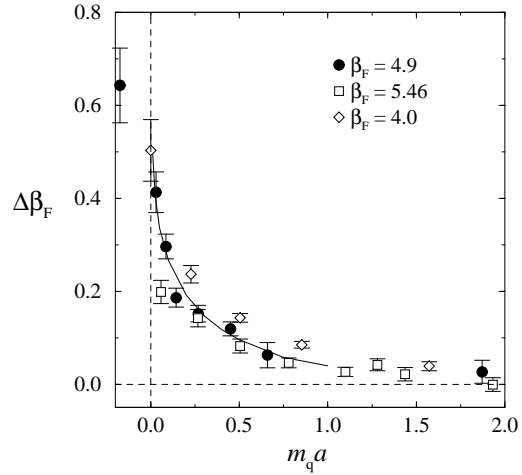

Figure 4. Comparison of the analytical result for $\Delta\beta_F$ [7] by Hasenfratz and DeGrand (solid line) to the MC data.

When $\kappa$ is small, the quarks are very massive and the induced couplings are quite small (left side of figs. 1 and 2). When $\kappa$ is increased, we approach the critical line where $m_q \to 0$, and the fermionic contribution to the action becomes more significant. This is clearly visible as an increase in $\beta_F$ in figs. 1 and 3. The critical values of $\kappa$ are approximately 0.16 ($\beta_F^0 = 5.46$), 0.19 (4.9) and 0.22 (4.0). We observe no significant increase in $\beta_A$; on the contrary, when $\kappa_c$ is approached, induced $\beta_A$ becomes slightly negative! The minor role of the adjoint action is also evident from the fact that $\beta_F$ remains practically the same whether we use the $S_A$-term of the effective action or not. We also checked the results with a few simulations on $6^4$ lattices with similar results.

In fig. 4 we compare $\Delta\beta_F$ to the predictions of ref. [7], as a function of quark mass $m_q a = \kappa^{-1} - \kappa_c^{-1}$. The agreement is very good, especially when $\beta_F^0 = 4.9$.

To conclude with, we did not observe that Wilson fermions induce any significant adjoint pure gauge action, and that the induced fundamental action is very well described by analytical calculations. It is very improbable that the transition observed in $N_t = 6$ Wilson thermodynamical simulations [2] could be explained by the fundamental-adjoint pure gauge transition, and the real cause of this phenomenon remains to be uncovered.

We thank the Institute for Theoretical Physics, Santa Barbara, where this project was initiated. K. R. would like to thank M. Hasenbusch for many useful discussions. This research was supported by NSF and DOE grants.